# Carbon Star CGCS 673 identified as a Semi-regular variable star

Stephen M. Brincat[1], C. Galdies[2] and K. Hills[3]

[1] Flarestar Observatory (MPC 171), Fl. 5/B, George Tayar Street, San Gwann, SGN 3160, Malta; *stephenbrincat@gmail.com*
[2] Institute of Earth Systems, University of Malta
[3] Tacande Observatory, El Paso, La Palma, Spain

**Abstract** This study shows that the carbon star CGCS 673 is a semi-regular (SR) variable star with a period of 135 d and an amplitude of 0.18 magnitudes in the V-band. The light curve obtained by this study correlates well with the SR classification as the photometric data obtained shows noticeable periodicity in the light changes of CGCS 673 that is occasionally interrupted by a period of irregular variability. The derived period and colour index obtained from our data and those from professional databases indicate that the attributes of this star fall within the parameters of the Semi-Regular class of variable stars. Following our notification of the discovery that this star is a variable source, CGCS 673 has received the AAVSO Unique Identifier as (AAVSO UID) 000-BMZ-492.

## 1 INTRODUCTION

Carbon stars share some of the characteristics of giant stars. Both reside at a late phase of stellar evolution. What makes a carbon star different from their more common counterparts is the fact that the ratio of carbon to oxygen ratio in their atmosphere is much higher. Hence the appropriate name given. Carbon stars reside within the Asymptotic Giant Branch (ABG) of the H-R diagram and their placement indicates that these old stars are in a phase that preceded the ejection of their atmospheres to form a planetary nebula. This phase of their lives has long been of interest to astronomers. Alksnis et al. 2001 compiled a catalog of Galactic Carbon Stars (CGCS) that contained 6891 candidates. Nesci et al. 2018 reports that out of the CGCS population, 851 were classified as variable stars within the General Catalog of Variable Stars (GCVS, Samus et al. 2017, CDS B/gcvs) with around 45 percent of these that had a period of variability reported. The AAVSO VSX catalog (Watson et al. 2016) reports over 958 CGCS variables with reported periods, hence only a minority of carbon stars are known to be variable sources. Thanks to the monitoring of a suspected variable star UCAC4 690-029948 that resulted in a new vari-able star discovery by Brincat and Hills (VSX 2019), further investigation revealed that within the same field of view lies the Carbon Star GSC 0333-00416. Results of the data retrieved through data mining show that the carbon star CGCS 673 is also a variable source.

CGCS 673 (also designated as GSC 03333-00416) was first described as a carbon star by Nassau and Blanco 1954. They identified this star as CASE 72 (entry 72) in their list of carbon stars near the galactic equator. Stephenson 1973 also listed the star through his paper with a second revision published in 1989 (Stephenson 1989). Soyano and Maehara 1991 mentioned the star CGCS 673 through their findings on 226 cool carbon stars that were derived from V-band plates within the Perseus-Camelopardalis region.

Many stars in this list were identified by the authors as the same carbon stars in Stephenson 1989 general catalogue. Chen et al. 1993 also identified CGCS 673 within their candidate list. The General Catalog of Galactic Carbon stars by Stephenson (Stephenson 1989) was published through an updated and revised version by Alksnis et al. 2001 where information such as coordinates, magnitudes, spectral classification and references were updated.

## 2 OBSERVATIONS

CGCS 673 (C* 205 = CSI+47-04201 = 2MASS J04234199+4753050 = Case 72 = IRAS 04200+4746 UCAC2 47715951 = GSC 03333-00416 = Kiso C5-190 = UCAC3 276-65568 = UCAC4 690-029995 = AAVSO AUID 000-BMZ-492) is located at RA 04 23 41.996 and DEC +47 53 5.06 (J2000) in the constellation of Perseus.

The brightness readings of CGCS 673 through different catalogues ranging from the near Infrared to the optical region are shown in Table 1.

**Table 1** Brightness readings of CGCS 673.

| Catalogue | Photometric Bands | | Sources |
|---|---|---|---|
| 2MASS | J=6.814 0.02 | H = 5.450 0.02 | K = 4.856 0.02 |
| | | | Skrutskie et al. 2006 |
| APASS-DR9 | B = 16.935 0.09 | V = 13.412 0.03 | g = 15.131 0.11 |
| | | | Henden and Munari 2014 |
| GAIA DR2 | G = 11.1198 0.0036 | BP = 13.4479 0.0141 | RP = 9.7842 0.0061 |
| | | | Brown et al. 2018 |

The GAIA DR2 catalogue lists the star as '257953644153609344' (Brown et al. 2018) with a stellar effective temperature of about 3300K. The reddening at this location due to interstellar extinction is listed as E(EB-RB)mag = 2.25.

We discovered the variability of CGCS 673 during a study of the neighbouring object UCAC4 690-029948. A large colour index (CI) was uncovered through the use of V (551 nm) and I band (806 nm) passband. This finding prompted further research to reveal that CGCS 673 is a known carbon star.

As a number of carbon stars are know to be variable stars, a search was conducted through the ASAS-SN survey that can depict data through light curves (Kochanek et al. 2017). This search showed that the star is a variable source (Figure 1). The data downloaded covered a time space of 1444 days covering the period from December 1, 2014 (JD 2457007.88) to 29 November 2018 (JD 2458451.95). The ASAS-SN images utilised by this research were those taken through V and G bands. G-band pho-tometry has been transformed into V magnitudes through an offset that has been applied for all of our analysis.

The ASAS-SN survey obtained photometric magnitudes through differential aperture photometry with zero-points calibrated to the APASS catalog (Henden et al. 2016). The ASAS-SN survey initially utilised four 14 cm aperture Nikon telephoto lenses as telescopes. Each of the four telescopes utilize a 2048$^2$ ProLine back-illuminated CCD camera by Finger Lakes Instruments. ASAS-SN imaged the area of CGCS 673 in batches of two or three images per night with a frequency interval ranging from 1 to 5 days. Occasionally, a longer interval occurred due to unfavourable weather conditions.

In order to determine additional physical characteristics of this carbon star, we have gathered additional photometry through V (visual) band taken from the observatories listed in Table 2 and their contributions in Table 3.

Differential aperture photometry was used to obtain brightness readings. The magnitudes of comparison stars were taken from AAVSO, for the neighbouring star NSV 1536. The list of comparison stars shown in Table 4 were those used by all observers except for the ASAS-SN data that utilised the same APASS source but possibly different comparison stars.

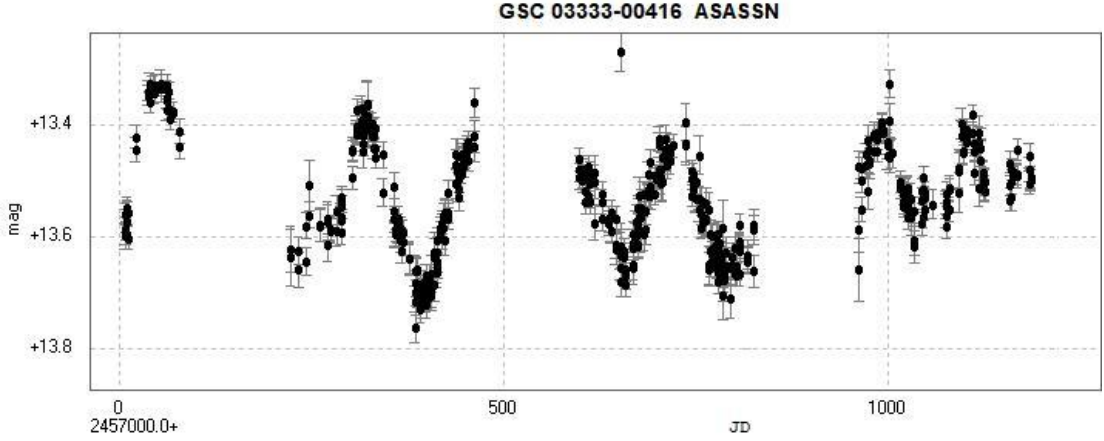

**Fig. 1** CGCS 673 light curve including ASAS-SN data. The light curve shows that the star is semi-regular in nature as regular cycles are interrupted by periods of irregularity.

**Table 2** Equipment used.

| Observatory (Location) | Observer | Telescope | Passband | CCD Sensor | FoV (arcmin)/Binning | Pixel Scale (arcsec/pixel) |
|---|---|---|---|---|---|---|
| Tacande Observatory (El Paso, La Palma, Spain) | Hills K. | 0.500-m ASA Astrograph | V | FLI ML3200/KAF 3200ME | 35.7x24.1 | 0.98 |
| Flarestar Observatory (San Gwann, Malta) | Brincat S. M. | 0.254-m SCT | V | Moravian G2-1600/KAF 1603 ME | 25.5x17.0/1x1 | 0.99 |
| Znith Observatory (Malta) | Galdies C. | 0.200-m SCT | V | Moravian G2-1600/KAF 1603 ME | 30.0x20.0/1x1 | 1.17 |

**Table 3** Observer contribution.

| Observer | Number of Observations | Band | Range Start (HJD) | Range End (HJD) | Span (d) |
|---|---|---|---|---|---|
| Brincat S.M. | 112 | V | 2458467 | 2458585 | 117.97 |
| Galdies C. | 9 | V | 2458549 | 2458573 | 24.04 |
| Hills K. | 227 | V | 2458405 | 2458505 | 100.78 |

## 3 DATA ANALYSIS

In order to investigate the periodic behaviour of CGCS 673, we performed a period search using Fourier analysis through the software PERANSO - version 2.60 (Paunzen and Vanmunster 2016).

**Table 4** Comparison stars used for the observation of CGCS 673 that were based on the APASS catalogue.

| AUID | RA | DEC | Label | V | B-V |
|---|---|---|---|---|---|
| 000-BMT-664 | 04:23:35.54 | 47:58:43.0 | 142 | 14.215 | 0.659 |
|  | [65.89808655°] | [47.97861099°] |  | (-0.04) | (0.071) |
| 000-BMT-665 | 04:23:39.81 | 47:59:58.3 | 146 | 14.629 | 0.989 |
|  | [65.9158783°] | [47.99952698°] |  | (0.051) | (0.097) |
| 000-BMT-666 | 04:24:27.34 | 47:46:28.0 | 149 | 14.930 | 0.972 |
|  | [66.11391449°] | [47.77444458°] |  | (0.035) | (0.077) |

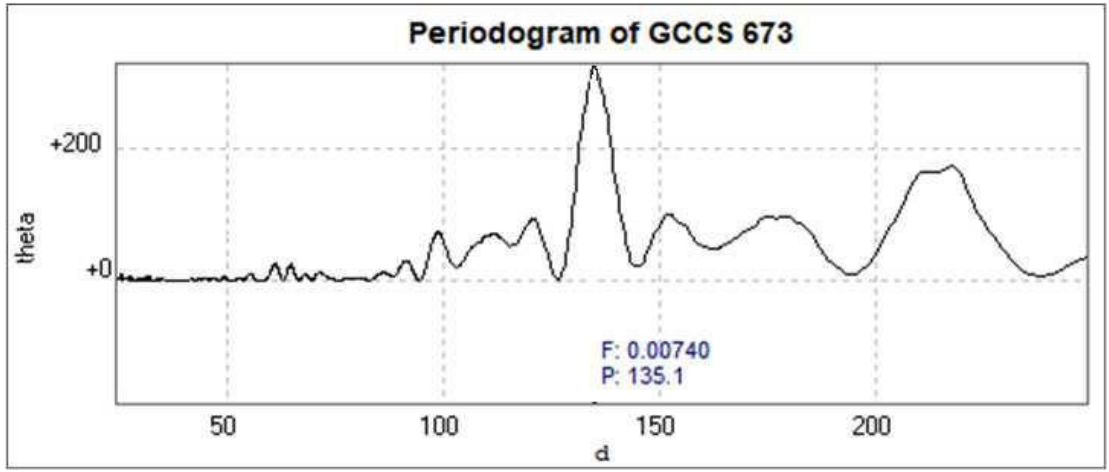

**Fig. 2** Periodogram of CGCS 673 as obtained through the Lomb-Scargle algorithm showing the peak at the derived period of 135.1 days (0.0074 c/d).

Period analysis was performed using the Phase Dispersion Minimization (PDM) algorithm (Stellingwerf 1978); ANOVA algorithm (Schwarzenberg-Czerny 1996) and Lomb-Scargle algorithm (Lomb 1976; Scargle 1982).

Through the algorithms mentioned above, the analysis of our observations along with those from the ASAS-SN Survey revealed the following periodicity within our dataset. The combined period of CGCS 673 was determined as 135.10 d 1.3 d with a mean amplitude (derived through a polynomial fit) in V-band of 0.188 mag (Figure 3). Figure 2 shows the periodogram of the Lomb-Scargle algorithm that shows identical results as obtained through Phase Dispersion Minimisation (PDM) algorithm and ANOVA algorithm.

## 4 CONCLUSIONS

Our observational campaign to monitor CGCS 763 was concluded on 11 April 2019, yielding a total number of 348 observations gathered by the observatories shown in Tables 3 and 4 for which several cycles were recorded. Our analysis of ASAS-SN data combined with those of our own observations show that the carbon star - GCCS 673 is a semi-regular (SR) variable star with a period of 135 d with amplitude of 0.18 magnitudes in the V-band. The light curve obtained by this study correlates well with the SR classification as the photometric data obtained shows the classical S-R morphological features in the changes of CGCS 673 that are occasionally interrupted by a period of irregular variability. The derived period, and colour index obtained from professional databases indicate that the attributes of this star fall within the parameters of the Semi-Regular class of variable stars.

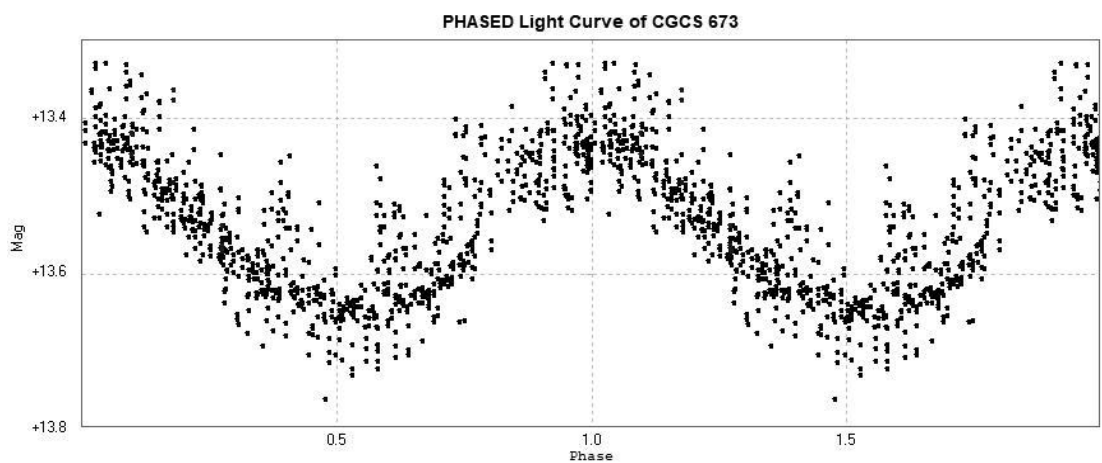

**Fig. 3** Phased lightcurve of CGCS 673 (Epoch: HJD 2457050.369) with data shown from the ASAS-SN Survey and data obtained by the observers listed in Table 3.

Following our notification that this star is a variable source, this star has received the AAVSO Unique Identifier as (AAVSO UID) 000-BMZ-492.

**Acknowledgements** This research was made possible through the use of the AAVSO Photometric All-Sky Survey (APASS), funded by the Robert Martin Ayers Sciences Fund and NSF AST-1412587.